\documentclass[fleqn,12pt,twoside]{article}
\usepackage{espcrc1,epsfig}

\usepackage{graphicx}
\usepackage[figuresright]{rotating}

\newcommand{\be}{\begin{equation}}
\newcommand{\ee}{\end{equation}}
\newcommand{\ba}{\begin{eqnarray}}
\newcommand{\ea}{\end{eqnarray}}

\newcommand{\AmS}{{\protect\the\textfont2
  A\kern-.1667em\lower.5ex\hbox{M}\kern-.125emS}}

\title{Dynamical Generation of Hyperon Resonances}

\author{A. Ramos,\address[MCSD]{Departament d'Estructura i Constituents de la Mat\`eria,
Universitat de Barcelona, \\ 
        Diagonal 647, 08028 Barcelona, Spain}%
        E. Oset,\address {Departamento de F\'{\i}sica Te\'orica and IFIC,
Centro Mixto Universidad de Valencia-CSIC,\\
 Institutos de Investigaci\'on de Paterna, Aptd. 22085, 46071 Valencia, Spain}
 {\bf C.} Bennhold, \address {Center for Nuclear Studies, Department of Physics,\\
 The George Washington University, Washington D.C. 20052 , USA}
  D. Jido,\address {ECT$^*$, Villa Tambosi, Strada delle Tabarelle 286, I-38050
  Villazano (Trento), Italy}\\
  J.A. Oller \address {Departamento de F\'{\i}sica, Universidad de Murcia, 30071
  Murcia, Spain}
  and 
  U.-G. Mei{\ss}ner\address {Universit\"at Bonn, Helmholtz-Institut f\"ur Strahlen- und
Kernphysik (Theorie)\\
 Nu{\ss}alle 14-16, D-53115 Bonn, Germany}
\address{ FZ J\"ulich, IKP (TH), D-52425 J\"ulich, Germany}
}
 
\begin{document}

\maketitle

\begin{abstract}
 In this talk we report on how, using a chiral unitary approach for the 
 meson--baryon interactions, two octets of $J^{\pi}=1/2^-$  baryon states
  and a singlet are generated dynamically,  resulting in the case of
strangeness $S=-1$ in two poles of the scattering matrix close to the nominal
$\Lambda(1405)$ resonance.   We  suggest experiments which could show
evidence for the existence of these states.
\end{abstract}

\section{Introduction}
The introduction of unitarity constraints in coupled channels in chiral
perturbation theory has led to unitary extensions of the theory that starting
from the same effective Lagrangians allow one to make predictions at much higher
energies. One of the interesting
consequences of these extensions is that they generate dynamically low lying
resonances, both in the mesonic and baryonic sectors. By this we mean that they
are generated by the multiple scattering of the meson or baryon components,
much  the same as the deuteron is generated by the interaction of the nucleons
through the action of a  potential, and they are not preexistent states that
remain in the large $N_c$ limit where the multiple scattering is suppressed. In
what follows we show how two octets and a singlet  of such states are generated
in the baryon sector with $J^P=1/2^-$, with properties in good agreement with
existing resonances. In the strangeness $S=-1$ case the lowest lying of such
resonances is the $\Lambda(1405)$.
The $\Lambda(1405)$ resonance is a clear example of a dynamically
generated resonance appearing naturally in scattering theory with coupled 
meson--baryon channels with strangeness $S=-1$ \cite{dalitz}. Chiral formulations
of the meson--baryon interaction within unitary frameworks all lead to the
generation of this resonance, which is seen  in the
 mass distribution of $\pi \Sigma$ states with isospin $I=0$ in hadronic production processes
 \cite{norbert,angels,oller,phase}. Yet, it was shown that in some models one
could obtain two poles close to the nominal $\Lambda(1405)$
resonance, as it was the case within the cloudy bag model in
Ref.~\cite{fink}. Also, in the investigation of the poles of the
 scattering matrix in Ref.~\cite{oller}, within the
 context of chiral dynamics, it was found that
 there were two poles close to the nominal $\Lambda(1405)$ resonance both
 contributing to the $\pi\Sigma$ invariant mass distribution.  This
 was also the case in Refs.~\cite{magnetic,nieves}, where
 two poles are obtained with similar properties as to their masses, widths and
 partial decay widths compared to those of the previous works. 

 The discussion below will clarify this issue and the structure of these
 resonances.

\section{Description of the meson baryon interactions}
\label{sec:2}
    Starting from the chiral Lagrangians for meson--baryon interactions
\cite{chiral} and using the N/D method to obtain a scattering matrix fulfilling
exactly unitarity in coupled channels \cite{oller}, the full set of transition matrix
elements with the coupled channels in  $S=-1$, $K^- p$,
 $\bar{K}^0 n$, $\pi^0 \Lambda$, $\pi^0 \Sigma^0$,
$\pi^+ \Sigma^-$, $\pi^- \Sigma^+$, $\eta \Lambda$, $\eta
\Sigma^0$, $K^0\Xi^0$ and
$K^+\Xi^-$,  is given in matrix form by

\begin{equation}
T = [1 - V \, G]^{-1}\, V~.
\label{eq:bs1}
\end{equation}
Here, the matrix $V$, obtained from the lowest order meson--baryon
chiral Lagrangian, contains the Weinberg-Tomozawa or seagull
contribution, as employed e.g.  in Ref.~\cite{bennhold},

\begin{equation}
V_{i j} = - C_{i j} \frac{1}{4 f^2}(2\sqrt{s} - M_{i}-M_{j})
\left(\frac{M_{i}+E}{2M_{i}}\right)^{1/2} \left(\frac{M_{j}+E^{\prime}}{2M_{j}}
\right)^{1/2}~ ,
\label{eq:ampl2}
\end{equation}
where the $C_{i j}$ coefficients are given in Ref.~\cite{angels},
and an averaged meson decay constant $f=1.123f_{\pi}$ is used
\cite{bennhold}, with $f_\pi = 92.4\,$MeV the weak pion decay constant.
 At lowest order in the chiral expansion all the
baryon masses are equal to the one in the chiral limit, $M_0$,
nevertheless in Ref.~\cite{bennhold} the physical baryon masses,
$M_i$, were used and these are the ones appearing in
Eq.~(\ref{eq:ampl2}). In addition to the Weinberg-Tomozawa term,
one also has at the same order in the chiral expansion the direct
and exchange diagrams considered in Ref.~\cite{oller}. These are
suppressed at low energies by powers of the three-momenta and
meson masses over $M_{i}$, the leading one being just linear.
However, their importance increases with energy and around
$\sqrt{s}\simeq 1.5$ GeV they can be as large as a $20$ percent of the
seagull term.

The diagonal
matrix $G$ stands
for the loop function of a meson and a baryon and is defined by a subtracted
 dispersion
relation in terms of phase space with a cut starting at the corresponding
threshold \cite{oller}. It
corresponds to the loop function of a meson and a baryon once the
logarithmic divergent constant is removed.

The analytical properties of $G$ are properly kept when evaluating
the previous loop function in dimensional regularization.
Using dimensional regularization and removing the divergent constant piece leads to
\begin{eqnarray}
G_{l} &=& i \, 2 M_l \int \frac{d^4 q}{(2 \pi)^4} \,
\frac{1}{(P-q)^2 - M_l^2 + i \epsilon} \, \frac{1}{q^2 - m^2_l + i
\epsilon}  \nonumber \\ &=& \frac{2 M_l}{16 \pi^2} \left\{ a_l(\mu) + \ln
\frac{M_l^2}{\mu^2} + \frac{m_l^2-M_l^2 + s}{2s} \ln \frac{m_l^2}{M_l^2} +
\right. \nonumber \\ & &  \phantom{\frac{2 M}{16 \pi^2}} +
\frac{q_l}{\sqrt{s}}
\left[
\ln(s-(M_l^2-m_l^2)+2 q_l\sqrt{s})+
\ln(s+(M_l^2-m_l^2)+2 q_l\sqrt{s}) \right. \nonumber  \\
& & \left. \phantom{\frac{2 M}{16 \pi^2} +
\frac{q_l}{\sqrt{s}}}
\left. \hspace*{-0.3cm}- \ln(-s+(M_l^2-m_l^2)+2 q_l\sqrt{s})-
\ln(-s-(M_l^2-m_l^2)+2 q_l\sqrt{s}) \right]
\right\}~ ,
\label{eq:gpropdr}
\end{eqnarray}
where $\mu$ is the scale of dimensional regularization. For a given value of this scale, the
subtraction constant $a_i (\mu) $ is determined so that the results are
finally scale independent.

This meson baryon loop function  was calculated
in Ref.~\cite{angels} with a cut-off regularization, similarly as
previously done in meson--meson scattering \cite{npa}.
%
The values of the $a_{i}$
 constants in Eq.~(\ref{eq:gpropdr})
are found to be around $-2$ to agree with the results of
the cut--off method for cut--off values of the order of the mass of
the $\rho(770)$ \cite{oller}, which we call of natural size.
Indeed, in Ref.~\cite{bennhold} it was found that with the values
for the subtraction constants 
\begin{equation}
\begin{array}{lll} a_{{\bar K}N}=-1.84~{\mathbf ,} &
a_{\pi\Sigma}=-2.00~,~ & a_{\pi\Lambda}\,=-1.83~, \\ a_{\eta
\Lambda}\,\,=-2.25~,~ & a_{\eta\Sigma}=-2.38~,~ & a_{K\Xi}=-2.67 \ ,
\end{array}
\label{eq:coef}
\end{equation}
one reproduces the results for the $G$ functions obtained in
Ref.~\cite{angels} with a cut--off of 630 MeV.

\section{Poles of the T-matrix}
\label{sec:3}

  The study of Ref.~\cite{bennhold} showed the presence of  poles in
  Eq.~(\ref{eq:bs1}) around the $\Lambda(1405)$ and the
$\Lambda(1670)$ for isospin $I=0$ and around the $\Sigma(1620)$ in
$I=1$.  The same approach in $S=-2$ leads to the resonance
$\Xi(1620)$ \cite{xi} and in $S=0$ to the $N^*(1535)$
\cite{siegel,inoue}, this latter one also generated dynamically in
Ref.~\cite{siegel}.
 One is thus tempted to consider the appearance of a singlet and an octet
of meson--baryon resonances. Nevertheless, the situation is more
complicated because indeed in the SU(3) limit there are {\it two} octets
and not just one, as we discuss below. 
The presence of these multiplets was already
  discussed in Ref.~\cite{oller} after obtaining a pole with $S=-1$ in the
$I=1$ channel,
with mass around 1430 MeV, and two poles with $I=0$, of masses around that of the
  $\Lambda(1405)$.
  Similar ideas have been exploited in the meson--meson interaction where a
nonet of dynamically generated mesons, made of  the $\sigma(500)$, $f_0(980)$,
$a_0(980)$ and $\kappa(900)$, has been obtained \cite{npa,ramonet,nsd,norbert2}.

 The appearance of a multiplet of dynamically generated mesons and baryons seems
 most natural once a state of the multiplet appears. Indeed, one must recall that the
 chiral Lagrangians are obtained from the combination of the octet of
 pseudoscalar mesons (the pions and partners) and the octet of stable baryons
(the nucleons and partners).  The SU(3) decomposition of the combination of two
 octets tells us that
 \begin{equation}
 8 \otimes 8=1\oplus 8_s \oplus 8_a \oplus 10 \oplus \overline{10} \oplus 27~.
\end{equation}
Thus, on pure SU(3) grounds, should we have a SU(3) symmetric Lagrangian,
 one can expect e.g. one singlet and two octets of resonances, the symmetric and
 antisymmetric ones.  Actually in the case of the meson--meson interactions only
 the symmetric octet appears in S-wave because of Bose statistics, 
but in the case of the meson--baryon
 interactions, where the building blocks come from two octets of different
 nature, both the symmetric and antisymmetric octets could appear and there is
 no reason why they should be degenerate in principle.

 The lowest order meson--baryon chiral Lagrangian is exactly SU(3) 
invariant if
 all the masses of the mesons 
are set equal. As stated above  [see Eq.~(\ref{eq:ampl2})], 
in Ref.~\cite{bennhold}
 the baryon masses take their physical values, although strictly
 speaking at the leading order in the chiral expansion they should be equal to
 $M_0$. For Eq.~(\ref{eq:ampl2}) being SU(3) symmetric, all the baryons masses
 $M_{i}$ must be set equal as well. When all the meson and baryon masses are
  equal, and these common masses are employed in evaluating the $G_l$ functions,
 together with equal subtraction constants $a_l$, the $T$--matrix obtained
from Eq.~(\ref{eq:bs1}) is also SU(3) symmetric. 

If we do such an SU(3) symmetry approximation
and look for poles of the scattering matrix, we find poles
corresponding to the octets and singlet. The surprising result is that
the two octet poles are degenerate as a consequence of the 
 dynamics contained in
 the chiral Lagrangians. Indeed, if we evaluate the matrix elements of the transition potential
$V$ in a basis of SU(3) states,

\begin{equation}
V_{\alpha  \beta}=\sum_{i,j} \langle i, \alpha \rangle C_{ij}
\langle j , \beta \rangle,
\end{equation}
where $\langle i, \alpha\rangle$ are the SU(3) Clebsch--Gordan
coefficients and $C_{ij}$ the coefficients in
Eq.~(\ref{eq:ampl2}), we obtain:

\begin{equation}
V_{\alpha  \beta}= {\rm diag}(6,3,3,0,0,-2)~, \label{eq:su3}
\end{equation}
taking the following order for the irreducible representations:
$1$, $8_s$, $8_a$, $10$, $\overline{10}$ and $27$.

Hence we observe that the states belonging to different
irreducible representations do not mix  and  the two octets appear
degenerate. The coefficients in
Eq.~(\ref{eq:su3}) clearly illustrate why there are no bound
states in the $10$, $\overline{10}$ and $27$ representations. 
Indeed, considering the minus sign in Eq.~(\ref{eq:ampl2}), a negative
sign in Eq.~(\ref{eq:su3}) means repulsion.

In practice, the same chiral Lagrangians allow for SU(3) breaking. In
the case of Refs.~\cite{angels,bennhold} the breaking appears
because both in the $V_{i j}$ transition potentials as in the $G_l$
loop functions one uses the
physical masses of the particles as well as different subtraction constants in $G_l$, 
corresponding to the use of a unique cut-off in all channels. 
 In Ref.~\cite{oller} the
physical masses are also used in the $G_l$ functions, although these functions are evaluated 
with a unique subtraction constant as corresponds to the SU(3) limit. In
addition, 
 the $V_{ij}$ transition potentials are evaluated strictly at lowest order in
 the chiral expansion, 
so that a common baryon mass is used and the one baryon exchange diagrams, both direct and crossed, 
are included. In both approaches, physical masses are
used to evaluate the $G_l$ loop functions so that unitarity is fulfilled
exactly and the physical thresholds for all channels are respected. 

\begin{figure}
  \epsfxsize=14cm
  \centerline{\epsfbox{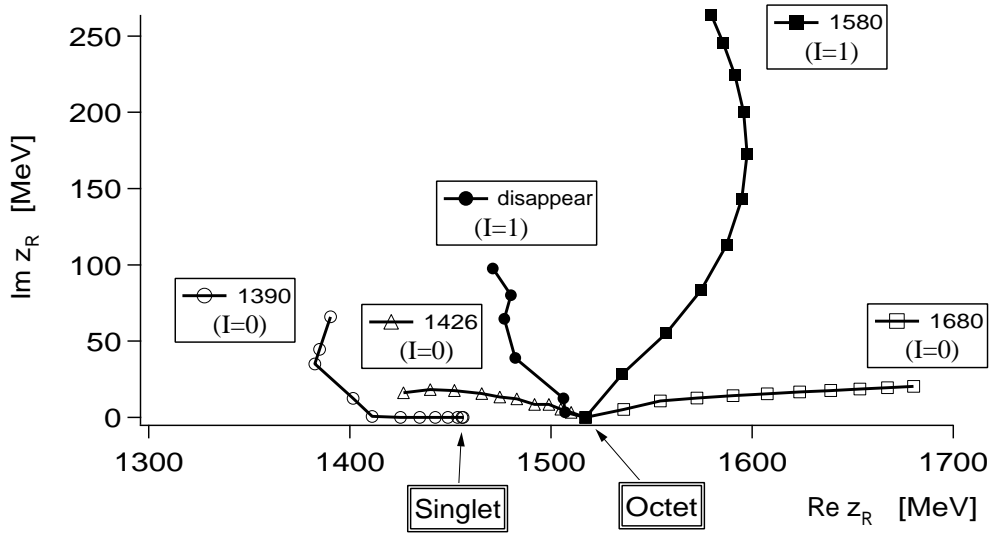}}
  \caption{Trajectories of the poles in the scattering amplitudes obtained by
  changing the SU(3) breaking parameter $x$ gradually. At the SU(3) symmetric 
  limit ($x=0$),
   only two poles positions appear, one for the singlet and the other for the octet.
  The symbols correspond to the step size $\delta x =0.1$. The results are from
  \cite{Jido:2003cb}.
  \label{fig:tracepole}}
\end{figure}

By following the approach of Ref.~\cite{bennhold} and using the
physical masses of the baryons and the mesons, the position of the
poles change and the two octets split apart in four branches, two
for $I=0$ and two for $I=1$, as one can see in
Fig.~\ref{fig:tracepole}. In the figure we show the trajectories
of the poles as a function of a parameter $x$ that breaks
gradually the SU(3) symmetry up to the physical values.  The
dependence of masses and subtraction constants on the parameter
$x$ is given by 
\begin{equation}
M_i(x) = M_0+x(M_i-M_0),  \quad
m^{2}_{i}(x) = m_{0}^{2} + x (m^{2}_{i}-m^{2}_{0}), \quad
a_{i}(x) = a_{0} + x (a_{i} - a_{0}),
\end{equation}
where $0\le x \le 1$. For the baryon masses, $M_{i}(x)$, the
breaking of the SU(3) symmetry follows linearly, while for the
meson masses, $m_{i}(x)$, the law is quadratic in the masses,
since in the QCD Lagrangian the flavor SU(3) breaking appears in
the quark mass terms and the squares of the meson masses depend on
the quark masses linearly.  In the calculation of
Fig.~\ref{fig:tracepole}, the values $M_{0}=1151$ MeV, $m_{0} =
368$ MeV and $a_{0}= -2.148 $ are used.

  The complex poles, $z_R$, appear in unphysical sheets. In the
present search we follow the strategy of changing
the sign of the momentum $q_l$ 
in the $G_l(z)$ loop function of
  Eq.~(\ref{eq:gpropdr}) for the channels which
are open at an energy equal to Re($z$).

The splitting of the two $I=0$ octet states is very interesting.
One moves to higher energies giving rise to the $\Lambda(1670)$
resonance and the other one moves to lower energies to create a
pole, quite well identified below the  $\bar{K}N$ threshold, with
a narrow width. 
We should also note that when for some values of
$x$ the trajectory crosses the $\bar{K}N$ threshold ($\sim 1435$
MeV) the pole fades away but it emerges again clearly for values
of $x$ close to $1$. On the other hand, the singlet also evolves
to produce a pole at low energies with a quite large width.

We note that the singlet and the $I=0$ octet states appear
nearby in energy and one of the purposes of this paper is,
precisely, to point out the fact that what experiments  actually
see is a combination of the effect of these two resonances.

Similarly as for the $I=0$ octet states, we can see that one branch of the 
$I=1$ states moves to higher energies while another
moves to lower energies. The branch moving to higher energies finishes at
what would correspond to the $\Sigma(1620)$ resonance when the physical
masses are reached. The
branch moving to lower energies fades away after a while when getting close to
the $\bar{K}N$ threshold.  

The model of Ref.~\cite{oller} reproduces qualitatively the same results.
However, this model also produces 
in the physical limit ($x=1$) another $I=1$ pole having Re($z$)=1401~MeV if,
in addition to changing the signs of
the on-shell momenta in the $\pi\Lambda$ and $\pi\Sigma$ channels
in accordance to the strategy mentioned above, the sign in the
${\bar K} N$ channel is also changed. The fact is that in both approaches there
is a $I=1$ amplitude with an enhanced strength around the $\bar{K} N$ thhrehold.
 Whether this enhancement in the $I=1$ amplitude
can be interpreted as a 
resonance or as a cusp, 
the fact that the
strength of the $I=1$ amplitude around the
$\Lambda(1405)$ region is not negligible should have consequences for 
reactions producing $\pi \Sigma$ pairs in that region.
This has been illustrated for instance
in Ref.~\cite{nacher1}, where the photoproduction of the $\Lambda(1405)$ via the
reaction $\gamma p \to K^+ \Lambda(1405)$ was studied. It was shown there that
the different sign in the
$I=1$ component of the $\mid \pi^+ \Sigma^-\rangle$, $\mid \pi^- \Sigma^+\rangle$
states leads, through interference between the $I=1$ and the dominant $I=0$
amplitudes, to different
cross sections in the various charge channels, a fact that has been
confirmed experimentally very recently \cite{ahn}.

Once the pole positions are found, one can also determine the
couplings of these resonances to the physical states by studying
the amplitudes close to the pole and identifying them with
\begin{equation}
T_{i j}=\frac{g_i g_j}{z-z_R}~.
\end{equation}
The couplings $g_i$ are in general complex valued numbers.
In Table~\ref{tab:jido0}  we summarize the
pole positions and the complex couplings $g_i$ obtained from the
model of Ref.~\cite{bennhold} for isospin $I=0$. The results with the model of
\cite{oller} are qualitatively similar.

\begin{table}[ht]
\centering \caption{\small Pole positions and couplings to $I=0$
physical states from the model of Ref.~\cite{bennhold}}
 \vspace{0.5cm}
\begin{tabular}{|c|cc|cc|cc|}
\hline
 $z_{R}$ & \multicolumn{2}{c|}{$1390 + 66i$} &
\multicolumn{2}{c|}{$1426 + 16i$} &
 \multicolumn{2}{c|}{$1680 + 20i$}  \\
 $(I=0)$ & $g_i$ & $|g_i|$ & $g_i$ & $|g_i|$ & $g_i$ & $|g_i|$ \\
 \hline
 $\pi \Sigma$ & $-2.5-1.5i$ & 2.9 & $0.42-1.4i$ & 1.5 & $-0.003-0.27i$ &
 0.27 \\
 ${\bar K} N$ & $1.2+1.7i$ & 2.1 & $-2.5+0.94i$ & 2.7 & $0.30+0.71i$ &
 0.77 \\
 $\eta\Lambda$ & $0.010+0.77i$ & 0.77 & $-1.4+0.21i$ & 1.4 & $-1.1-0.12i$ &
 1.1 \\
 $K\Xi$ & $-0.45-0.41i$ & 0.61 & $0.11-0.33i$ & 0.35 & $3.4+0.14i$ &
 3.5 \\
 \hline
 \end{tabular}
\label{tab:jido0}
\end{table}

\begin{table}[ht]
\centering \caption{\small Pole positions and couplings to $I=1$
physical states from the model of Ref.~\cite{oller}}
\vspace{0.5cm}
\begin{tabular}{|c|cc|cc|}
  \hline
$z_{R}$ & \multicolumn{2}{c|}{$1401+40i$} &
\multicolumn{2}{c|}{$1488+114i$} \\
 $(I=1)$ & $g_i$ & $|g_i|$ & $g_i$ & $|g_i|$  \\
 \hline
 $\pi\Lambda$ & $0.60+0.47 i$ & 0.76 & $0.98+0.84 i$ & 1.3 \\
 $\pi \Sigma$ & $1.27+0.71 i$ & 1.5 & $-1.32-1.00 i$ & 1.7 \\
 ${\bar K} N$ & $-1.24-0.73 i$ & 1.4 & $-0.89-0.57 i$ & 1.1 \\
 $\eta\Sigma$ & $0.56+0.41 i$ & 0.69 & $0.58+0.29 i$ & 0.65 \\
 $K\Xi$ & $0.12+0.05i$ & 0.13 & $-1.63-0.91i$ & 1.9 \\
  \hline
\end{tabular}
\label{tab:oller1}
\end{table}

We now consider the results obtained from the model of
Ref.~\cite{oller}. Making use of their set~I of parameters, which
correspond to a  baryon mass $M_0=1286$ MeV and a meson decay
constant $f=0.798f_{\pi}=74.1$ MeV, both in the chiral limit,
together with a common subtraction constant $a=-2.23$, the results
obtained for  $I=1$ are
displayed in Table~\ref{tab:oller1}.

 We observe that the
second resonance with $I=0$ couples strongly to $\bar{K} N$ channel, while
the first resonance couples more strongly to $\pi \Sigma$. The
results for $I=0$ shown in Table~\ref{tab:jido0}
resemble much
those obtained in Ref.~\cite{fink} and Ref.~\cite{nieves} where
two resonances are also found close to 1405~MeV, with the one at
lower energies having a larger width than the second and a
stronger coupling to $\pi\Sigma$, while the resonance at higher
energies being narrower and coupling mostly to $\bar{K}N$.

We can also project the states found over the pure SU(3) states and we find the
results of Table~\ref{tab:su30}

\begin{table}[ht]
\centering \caption{\small Couplings of the $I=0$ bound states to
the meson--baryon SU(3) basis states, obtained with the model of
Ref.~\cite{bennhold}} \vspace{0.5cm}
\begin{tabular}{|c|cc|cc|cc|}
\hline
 $z_{R}$ & \multicolumn{2}{c|}{$1390 + 66i$} &
\multicolumn{2}{c|}{$1426 + 16i$} &
 \multicolumn{2}{c|}{$1680 + 20i$} \\
  & \multicolumn{2}{c|}{(evolved singlet)} &
\multicolumn{2}{c|}{(evolved octet $8_s$)} &
 \multicolumn{2}{c|}{(evolved octet $8_a$)} \\
 & $g_\gamma$ & $|g_\gamma|$ & $g_\gamma$ & $|g_\gamma|$ & $g_\gamma$ & $|g_\gamma|$ \\
 \hline
 1 & $2.3+2.3i$ & 3.3 & $-2.1+1.6i$ & 2.6 & $-1.9+0.42i$ &
 2.0 \\
 $8_s$ & $-1.4-0.14i$ & 1.4 & $-1.1-0.62i$ & 1.3 & $-1.5-0.066i$ &
 1.5 \\
 $8_a$ & $0.53+0.94i$ & 1.1 & $-1.7+0.43i$ & 1.8 & $2.6+0.59i$ &
 2.7 \\
 $27$ & $0.25-0.031i$ & 0.25 & $0.18+0.092i$ & 0.21 & $-0.36+0.28i$ &
 0.4 \\
 \hline
 \end{tabular}
\label{tab:su30}
\end{table}

We observe that the physical singlet couples mostly to the singlet
SU(3) state. This means that this physical state has retained
largely the singlet nature it had in the SU(3) symmetric
situation. The same is true for the physical $I=0$  antisymmetric
octet shown in the last column. However, the couplings of the
physical symmetric octet reveal that, due to its proximity to the
singlet state, it has become mostly a singlet with some admixture
of the symmetric and antisymmetric octets.

\section{Influence of the poles on the physical observables}
\label{sec:5}

\begin{figure}
\epsfxsize=12cm
\centerline{\epsfbox{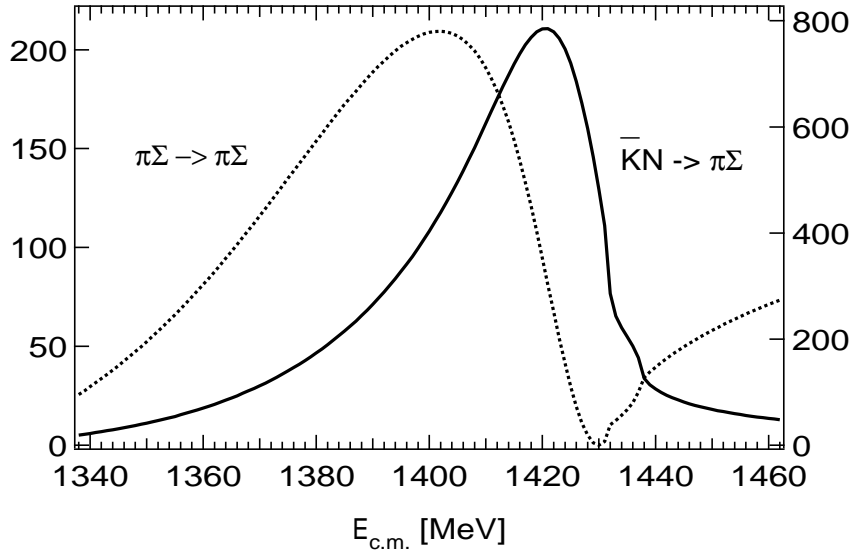}}
\caption{The $\pi\Sigma$ mass distributions with $I=0$ constructed from
  the $\bar{K}N\rightarrow \pi\Sigma$ and $\pi\Sigma \rightarrow \pi\Sigma$
  amplitudes. The solid and dashed lines denote
  $|T_{\bar{K}N\rightarrow \pi\Sigma}|^{2} q_{\pi}$ and
  $|T_{\pi\Sigma\rightarrow \pi\Sigma}|^{2} q_{\pi}$, respectively.
  Units are arbitrary.
\label{fig:massdist}}
\end{figure}

In a given reaction the $\Lambda(1405)$ resonance is always seen in $\pi \Sigma$
mass distribution. However, this final state can be reached through the
production of any intermediate state which couples to the  $\pi \Sigma$ state,
since the final state interaction will reshufle the channels as we have seen in
the chiral unitary approach.  Hence, the mass distribution of   $\pi \Sigma$
will be given by

\begin{equation}
 {d \sigma \over dM_{i}} =| \sum_i C_i T_{i \rightarrow \pi \Sigma}|^{2}
     q_{\rm c.m.} \ ,
\label{eq:psn}
\end{equation}
where the $C_i$ coefficients depend on the dynamics of the particular problem.
This is curious since practically in all approaches where the $\Lambda(1405)$ is
claimed to be obtained one uses the equation

\begin{equation}
 {d \sigma \over dM_{i}} = C |T_{\pi \Sigma \rightarrow \pi \Sigma}|^{2}
     q_{\rm c.m.} \ ,
\label{eq:ps}
\end{equation}
where $C$ is a constant, which has no justification. Indeed, if the sum in
eq.~(\ref{eq:ps}) were dominated by the 
$\bar{K}N \rightarrow \pi\Sigma$
amplitude, then the second resonance $R_{2}$ would be weighted
more, since it has a stronger coupling to the $\bar{K}N$ state,
resulting into an apparent narrower resonance peaking at higher
energies. This can be seen in Fig.~\ref{fig:massdist}. In ref. \cite{oller} the
$C_i$ coefficients foor $\bar{K} N$ and $\pi \Sigma$ were fitted to the data.
The fact is that if there were only one resonance eqs.~(\ref{eq:psn}) and 
(\ref{eq:ps}) would lead to the same shape of the distribution since all the
amplitudes would have the same resonance shape. But the existence of two poles
makes the sum in eq.~(\ref{eq:psn}) dependent on the weights $C_i$ and then
dependent on the particular reaction. Hence, from now on, a theoretical claim 
about understanding the $\Lambda(1405)$ properties has to be substanciated by a
simultaneous theoretical analysis of the particular reaction where this resonance
has been seen.  In this sense, there is a recent work \cite{Hyodo:2003jw} in which the 
dynamics of
the $\pi^- p \to K^0 \pi \Sigma$, from where the nominal $\Lambda(1405)$ resonance
comes,  has been studied and the particular shape of the resonance found in this
reaction is traced back to a nontrivial combination of chiral mechanisms
involving the meson pole and contact term in the $MB \to MMB$ amplitudes
together with the contribution of the s-wave $N^*(1710)$ resonance which has a
strong coupling to the $MMB$ system, as proved by the large decay width into
$\pi \pi N$.

It is clear that,  should there be a reaction which forces the
initial channels to be $\bar{K}N$, then this would give more
weight to the second resonance, $R_{2}$, and hence produce a
distribution with a shape corresponding to an effective resonance
narrower than the nominal one and at higher energy. Such a case
indeed occurs in the reaction $K^- p \to \Lambda(1405) \gamma$
studied theoretically in Ref.~\cite{nacher}.  It was shown there
that since the $K^- p$ system has a larger energy than the
resonance, one has to lose energy emitting a photon prior to the
creation of the resonance and this is effectively done by the
Bremsstrahlung from the original $K^-$ or the proton.  Hence the
resonance is initiated from the  $K^- p$ channel.

\section{Conclusions}
\label{sec:6}

 In this talk we have shown the poles appearing in the
 meson--baryon scattering matrix for strangeness $S=-1$ within a 
coupled--channel chiral
unitary
 approach, using two different methods for breaking the SU(3)
symmetry which have been used in the literature.

   In both approaches a set of  resonances is generated dynamically from the
 interaction of the octet of pseudoscalar mesons with the octet of the
$1/2^+$
 baryons.  The underlying SU(3) structure of the Lagrangians implies
that, from
 the combination of the two original octets, a singlet and two octets of
 dynamically generated resonances should appear, but the dynamics of the
problem makes the two octets degenerate in the case of
exact SU(3) symmetry.  
The same chiral Lagrangians have mechanisms for  chiral
 symmetry breaking which have as a consequence that the degeneracy is broken
 and two distinct octets appear. 
  The breaking of the octet degeneracy has as a consequence that, in the
physical limit, one of the
  $I=0$ octet poles appears
 quite close to the singlet pole,
and both of them are
 very close to the nominal $\Lambda(1405)$.  These two resonances are quite
 close but different, the one at lower energies with a larger width and
a stronger 
 coupling to the $\pi \Sigma$ states than the one at higher energies, which
 couples mostly to the $\bar{K}N$ states. Thus we conclude that there is not just one single
$\Lambda(1405)$
 resonance, but {\em two}, and that what one sees in experiments is a
{\em superposition} of these two states.   
 
  Another interesting finding  is the suggestion that it is
 possible to find out the existence of the two resonances by performing
 different experiments, since in different experiments the weights by which the two
 resonances are excited are different.  In this respect we call the
attention to one reaction,  $K^-p \to \Lambda(1405) \gamma$, which gives much
 weight to the resonance which couples strongly to the $\bar{K}N$ states
and, hence, leads to a peak structure in the invariant mass distributions 
 which is
 narrower and appears at higher energies than the experimental
$\Lambda(1405)$
 peaks observed in hadronic experiments performed so far.
 
   Finally, the findings discussed here about a possible $I=1$ state close to
the $\bar{K} N$ threhold and the experiment done in \cite{ahn}, which shows very
distinct $\pi^+ \Sigma^-$ and $\pi^- \Sigma^+$ distributions, deserve further
attention and should lead in the near future to a clarification on the situation
of this hypothetical state.
                         
\section*{Acknowledgments}
This work is partially supported by DGICYT
projects BFM2000-1326, BFM2001-01868, FPA2002-03265,
the EU network EURIDICE contract
HPRN-CT-2002-00311,
and the Generalitat de Catalunya project
2001SGR00064.
D.J. would like to acknowledge the support of Japanese
Ministry of Education, Culture, Sports, Science and Technology
to stay at IFIC, University of Valencia, where part of this work was
done.

\end{document}